\documentclass[showpacs,preprintnumbers,amsmath,amssymb]{revtex4}

\usepackage{dcolumn}
\usepackage{bm}

\usepackage{epsf}

\def\c60{$\rm C_{60}$}
\def\nec60{Ne@\c60}
\def\hec60{He@\c60}
\def\kinc60{K@\c60}
\def\etal{{\it et al.}}
\def\kc60{$($K@\c60$)_n$}
\def\t1u{$t_{1u}$}
\def\t1g{$t_{1g}$}
\def\b1u{$b_{1u}$}
\def\b2u{$b_{2u}$}
\def\b3u{$b_{3u}$}
\def\b1g{$b_{1g}$}
\def\b2g{$b_{2g}$}
\def\b3g{$b_{3g}$}

\begin{document}
\title{Ferroelectricity in (K@C$_{60}$)$_n$}

\author{Dennis P. Clougherty\footnote{Permanent address: Department of Physics,
University of Vermont, Burlington, VT 05405-0125}}
\affiliation{Institute for Theoretical Atomic and Molecular Physics\\
Harvard-Smithsonian Center for Astrophysics\\
60 Garden St.\\
Cambridge, MA 02138}


\begin{abstract}
A theoretical analysis of the ground state of long-chain (K@C$_{60}$)$_n$ is
presented. Within mean field theory, a ferroelectric ground state
is found to be stable because of the pseudo-Jahn-Teller mixing of
the $b_{1u}$ and the
 $b_{2g}$ band with a zone-center optical phonon
involving the displacement of the endohedral {\rm K}$^+$ ions. A
phase diagram
 for this  model is derived in the narrow bandwidth regime.
\end{abstract}

\maketitle
\section*{Introduction}

Fullerenes have proven to be useful building blocks in the
synthesis of novel molecular solids. Metals, insulators,
ferromagnets, and superconductors have all been fashioned
out of fullerenes.  A variety of fullerene-based polymeric structures
 have also been synthesized and
characterized\cite{pekker,geckeler,rao,patil,stephens}.  Linear
chains of C$_{60}^-$ have been produced\cite{pekker} by the
coevaporation of K with C$_{60}$.  The external K atoms readily donate
their valence electrons to the C$_{60}$ molecules, forming a charged
``pearl necklace'' of C$_{60}$ anions (see Fig.~\ref{pearls}).

Endohedral doping, where one or more atoms are positioned inside
the C$_{60}$ cage, offers an alternative way to modify the properties
of fulleride solids. Theoretical studies show that sizeable
electric dipole
moments\cite{cioslowski,liu,dunlap,wang,erwin,clougherty} can
result from the endohedral doping of isolated fullerenes with
metal atoms, so-called metallofullerenes.  The dipole formation
can be understood\cite{clougherty} as a consequence of the
pseudo-Jahn-Teller effect (PJT) where the $t_{1u}$ and $t_{1g}$
electronic states of the metallofullerene are mixed by a $t_{1u}$
distortion involving the displacement of the central metal atom,
the $(t_{1u}\oplus t_{1g})\otimes t_{1u}$ model.

The properties of a linear chain of metallofullerenes are
investigated in what follows.  A lattice generalization of the
$(b_{1u}\oplus b_{2g})\otimes b_{3u}$ PJT model is explored. A
phase diagram
 for this  model is derived in the narrow bandwidth regime.  It is
shown that within mean field theory, a ferroelectric ground state
is stable when the coupling constant $g$ exceeds a critical value
$g_0$ which depends on the cation force constant $\kappa$
 and the $b_{1u}-b_{2g}$ energy spacing $\Delta$.

\section*{Model Hamiltonian}

\begin{figure}[b!] 
\epsfbox[100 125 575 375]{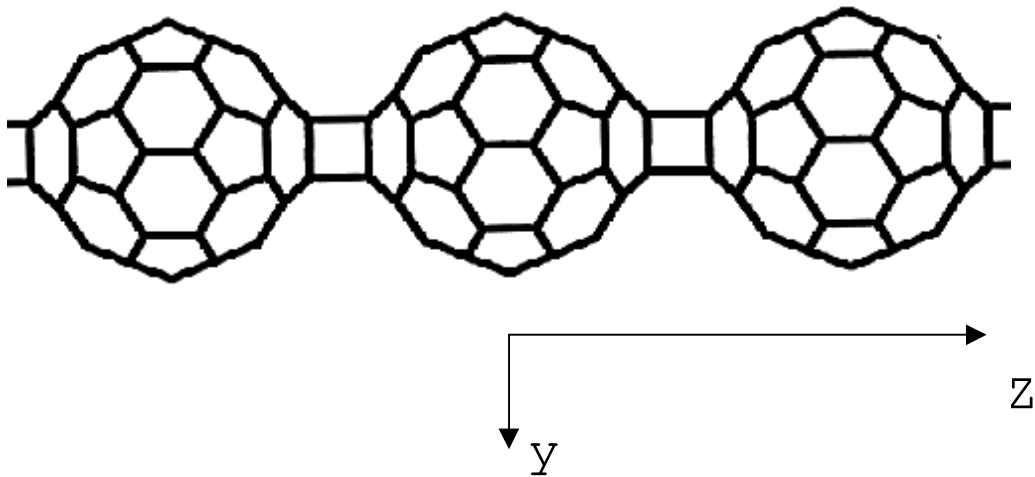} \caption{Bonding structure
of (KC$_{60}$)$_n$.} \label{pearls}
\end{figure}

ESR measurements on exohedral polymeric (KC$_{60}$)$_n$ are
consistent with one electron transfer from K atoms to the C$_{60}$
chains. Adjacent C$_{60}$ molecules bond with a preferred orientation
as indicated in Fig.~\ref{pearls}.  The axis of the chain bisects
a pair of hexagon-hexagon bonds on every C$_{60}$ molecule.  It is
assumed that endohedral polymeric (K@C$_{60}$)$_n$ will have the same
structure.

Such an arrangement of metallofullerenes with the K$^+$ ions
centered with respect to the C$_{60}$ cages will have a point group
symmetry of D$_{2h}$, as the chain axis is coincident with one of
the molecule's two-fold rotation axes. Facing hexagon-hexagon
bonds on adjacent C$_{60}$ molecules align, and the two intermolecular
bonds together with the two facing hexagon-hexagon bonds complete
a four-membered ring.  A coordinate system is chosen so that the
chain axis is labeled by $z$ and the four-membered rings lie in
the $yz$-plane as indicated in Fig.~\ref{pearls}.

In an isolated alkali-doped metallofullerene, the highest occupied
molecular orbital (HOMO) has a three-fold spatial degeneracy and
tranforms as t$_{1u}$ under the icosahedral group. With the site
symmetry lowered to D$_{2h}$, this three-fold degeneracy is lifted
and the t$_{1u}$ multiplet is split into its components under
D$_{2h}$: b$_{1u}$, b$_{2u}$, and b$_{3u}$. Ligand-field analysis
indicates that the b$_{2u}$ and b$_{3u}$ levels which transform as
$y$ and $x$ will be closely spaced in energy, while the b$_{1u}$
will be lowered by more than the splitting $\delta$ between
b$_{2u}$ and b$_{3u}$.

It can be argued using symmetry considerations that the b$_{2u}$
state will be lower in energy than the b$_{3u}$ state. The HOMO is
composed primarily of radially directed atomic $p$ orbitals. Since
the b$_{3u}$ state must transform as $x$, the $yz$-plane must be a
nodal plane for this state, as any combination of $p_z$ orbitals
centered on the sites of the four-membered ring would be even
under reflection through the $yz$-plane. Thus the C atoms
participating in the hexagon-hexagon bonds on axis can not
contribute $p_z$ character to the b$_{3u}$ orbital.

In the b$_{2u}$ state, however, these C atoms can contribute $p_z$
character and satisfy the symmetry requirement of oddness under
$y\to -y$. Consequently, the b$_{2u}$ state has a lower energy
through $\sigma$-bonding to nearest neighbors along the chain
axis.

\begin{figure}[] 
\epsfxsize=6in
\epsfbox[50 325 550 750]{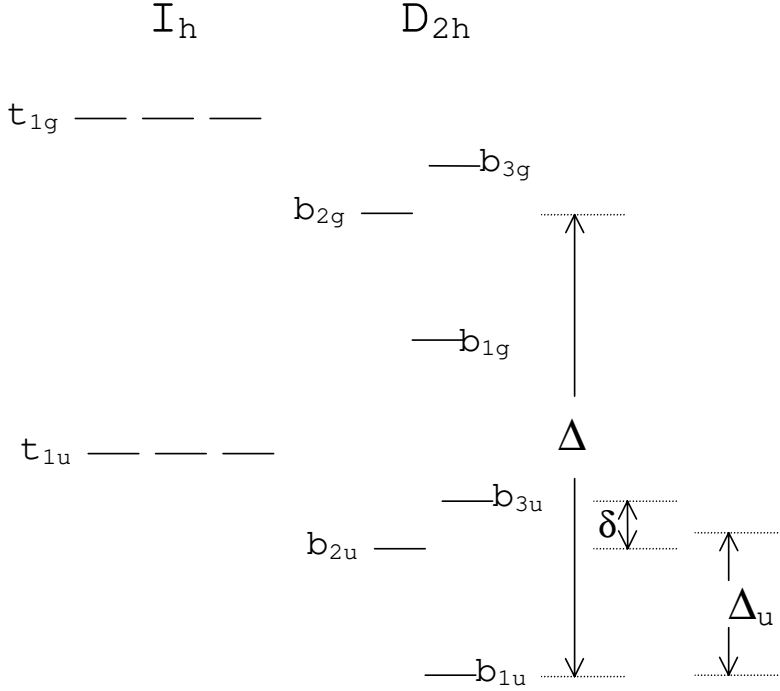}
\vspace{12pt}
\caption{Correlation diagram of t$_{1u}$ and t$_{1g}$ manifolds under an axial
ligand field.} \label{orbital}
\end{figure}

The lowest unoccupied molecular orbital (LUMO) in an alkali-doped
metallofullerene transforms as $t_{1g}$.  In analogy with the
HOMO, in the polymer, this multiplet is split into $b_{1g}$,
$b_{2g}$, and $b_{3g}$ with the ordering given in
Fig.~\ref{orbital}.


It has been previously shown\cite{clougherty} that there is strong
vibronic mixing of the $t_{1u}$ and $t_{1g}$ orbitals in
metallofullerenes through central atom displacements. A lattice
generalization of that pseudo-Jahn-Teller (PJT) model is now
described.


The Hamiltonian is given by
\begin{equation}
{\cal H}={\cal H}_{0}+{\cal H}_{{\rm PJT}}+{\cal
H}_{{\rm elast}}
\label{ham}
\end{equation}
where
\begin{equation}
{\cal H}_{0}=-t\sum_{n}  (c^\dagger_{1
n} c^{\phantom{\dagger}}_{1 n+1}+ c^\dagger_{2
n} c^{\phantom{\dagger}}_{2 n+1}+{\rm h.c.})+
\sum_{n}  \Delta c^\dagger_{2 n} c^{\phantom{\dagger}}_{2 n}
\label{band}
\end{equation}
where $n$ runs over endofullerene site, $c^{{\dagger}}_{1 n}$
($c^{\phantom{\dagger}}_{1 n}$) creates (annihilates) an electron
in orbital $b_{1u}$ at site $n$, $c^{{\dagger}}_{2 n}$
($c^{\phantom{\dagger}}_{2 n}$) creates (annihilates) an electron
in orbital $b_{2g}$ at site $n$, and $\Delta$ is the energy
spacing between the $b_{1u}$ and $b_{2g}$ orbitals. The hopping
matrix is taken to be diagonal in orbital type and spin, and the
spin indices are suppressed for clarity.

The pseudo-Jahn-Teller Hamiltonian on the $n$th endofullerene is
taken to be
\begin{equation}
{\cal H}_{{\rm PJT}}=  -{g\Delta\over a} \sum_{n} (c^\dagger_{1 n}
c^{\phantom{\dagger}}_{2 n} Q_{n}+{\rm h.c.}) \label{pjt}
\end{equation}
where $a$ is the equilibrium spacing between successive
endofullerenes, and $g$ is a dimensionless coupling constant.

 The point group symmetry only allows for vibronic mixing of
$b_{1u}$ and  $b_{2g}$ orbitals with a displacement transforming
as $b_{3u}$.  Hence, $Q_n$ is a displacement on the $n$th
endofullerene in the $x$ direction, perpendicular to the plane of
the four-membered rings.

${\cal H}_{{\rm elast}}$ is the elastic energy for the
displacements of the K$^+$ ions.
\begin{equation}
{\cal H}_{{\rm elast}}={\kappa\over 2}\sum_n Q_{n}^2
\label{elastic}
\end{equation}

With the following transformations,
\begin{equation}
c^{\phantom{\dagger}}_{\alpha k}={1\over\sqrt{N}}\sum_n {\rm
e}^{ikn} c_{\alpha n}
\end{equation}

\begin{equation}
Q_{k}={1\over\sqrt{N}}\sum_n {\rm e}^{ikn} Q_{n}
\end{equation}
the Hamiltonian can be written in the $k$-space representation
\begin{equation}
{\cal H}= \sum_{\alpha, k}{\epsilon_\alpha}(k) c^\dagger_{\alpha
k} c^{\phantom{\dagger}}_{\alpha k} -{g\Delta\over\sqrt{Na^2}}
\sum_{ k, q}  (c^\dagger_{1 k+q} c^{\phantom{\dagger}}_{2 k}
Q_{q}+{\rm h.c.}) + {1\over 2}\kappa\sum_{ q} |Q_{ q}|^2
\label{kham}
\end{equation}
where $\epsilon_{g}(k)=\Delta-2t\cos k$ and $\epsilon_{u}(k)=-2t\cos k$.

In analogy with (KC$_{60}$)$_n$, the t$_{1u}$ and t$_{1u}$ bands in
(K@C$_{60}$)$_n$ are expected to have a width of the order of $\frac{1}{2}$
eV. Hence, (K@C$_{60}$)$_n$ is expected to be in the narrow bandwidth regime
where the $b_{1u}-b_{2g}$ orbital splitting is much larger than
the bandwidth, $\Delta >> 4t$.

\section*{Mean Field Theory}

With a single mode of distortion given by a non-vanishing $Q_{q}$,
${\cal H}$ can be diagonalized with a canonical transformation so that
\begin{equation}
{\cal H}=\sum_k (E_{k+} n_{k+}+E_{k-} n_{k-})+
{\kappa\over 2}|Q_{q}|^2
\end{equation}
where
\begin{equation}
E_{k\pm}=\frac{1}{2}(\epsilon_u(k)+\epsilon_g(k+q))\pm \frac{1}{2}
\sqrt{(\epsilon_g(k+q)-\epsilon_u(k))^2+{4g^2\Delta^2\over
Na^2}|Q_q|^2}
\end{equation}
and $n_{k\pm}$ are the usual occupation number operators of the
vibronic quasiparticles.

\begin{figure}[t] 
\epsfxsize=6in
\epsfbox{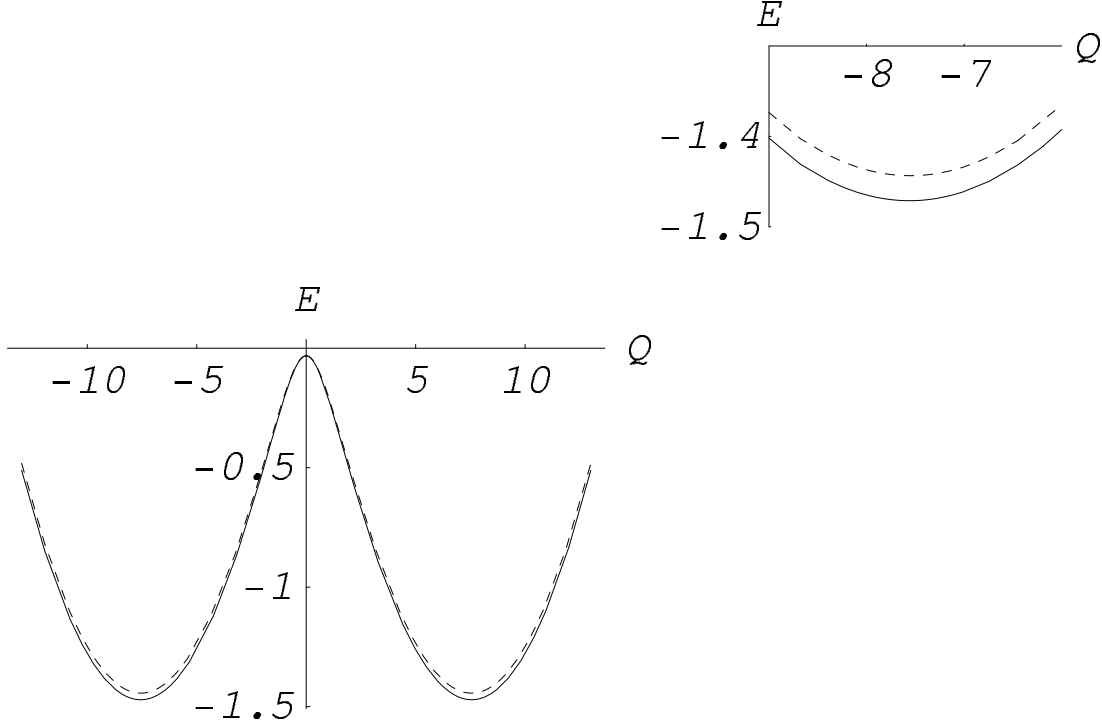}
\caption{Ground state energy
per unit length $E$ vs. $Q=2g\Delta |Q_q|/a\sqrt{N}$ for $q=0$ and
$q=\pi$ (dashed). The values $g=20$, $t=0.5$ eV, $\Delta=1$ eV and
$\kappa=1.6$ eV/\AA$^2$ were chosen. Top right graph shows detail
in vicinity of a minimum.}
\label{energy}
\end{figure}

The ground state energy of chain with one electron per endofullerene
on average is
obtained by half-filling the lowest energy vibronic band.
In the continuum limit,
\begin{equation}
E(Q_q)={Na\over \pi}\int_{-{\pi/2a}}^{\phantom{-}\pi/2a} dk\
E_{k-} +{\kappa\over 2}|Q_{q}|^2
\end{equation}

The ground state energy can be calculated to ${\cal O}\big(
{4t\over\Delta}\big)^2$
\begin{eqnarray}
E(Q_q)\approx  {Na\Delta\over 2}\bigg[&
1-\big({4t\over\Delta}\big){2\over\pi}\cos^2{qa/ 2}
-\sqrt{1+\tilde Q_q^2} -\big({4t\over\Delta}\big){\sin^2{qa/
2}\over\pi\sqrt{1+\tilde Q_q^2}}\cr &-\big({4t\over\Delta}\big)^2
{\tilde Q_q^2 \sin^2 {qa/2}\over 4({1+\tilde Q_q^2})^{3\over 2}}
(1+{1\over\pi}\sin qa)\bigg]+{\kappa}|Q_{q}|^2/2
\label{en}
\end{eqnarray}
where $\tilde Q_q^2={4 g^2 |Q_{q}|^2/ Na^2}$.

At $q=0$, the energy is a minimum for non-vanishing $Q_q$ when $g
> g_0$ where $g_0=\sqrt{\kappa a^2/ 2\Delta}$. The results for the
ferroelectric and antiferroelectric cases are displayed in
Fig.~\ref{energy}.

\begin{figure}[!] 
\epsfbox[0 0 288 177]{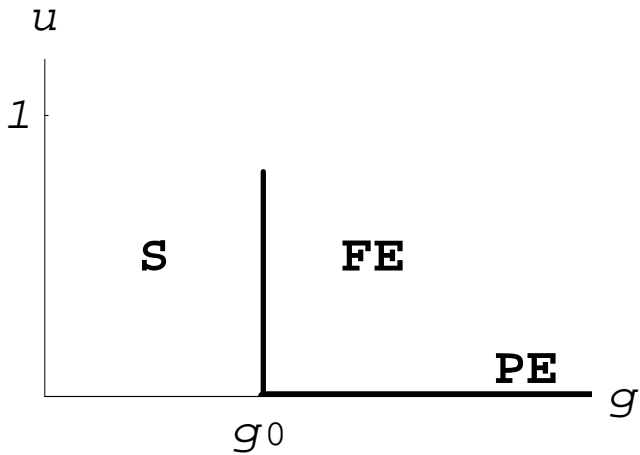}
\vspace{12pt}
\caption{Zero-temperature phase diagram for (K@C$_{60}$)$_n$ in
the narrow bandwidth regime $u={4t\over\Delta}<< 1$.}
\label{phase}
\end{figure}

When $g>g_0$ the energy minimum is found at $q=0$, corresponding
to an ferrodistortive transition. This displacement creates local
electric dipoles on the endofullerenes that are aligned
perpendicular to the $yz$-plane, yielding a uniaxial ferroelectric
phase (FE). A paraelectric (PE) phase exists along the line
segment $g > g_0$ and $t=0$ where independent, local dipole
moments form on the endofullerenes. The symmetric phase (S), where
no local dipole moments form, is stable for $g<g_0$. These results
are summarized in
 zero-temperature phase diagram pictured in Fig.~\ref{phase}.

\section*{Acknowledgments}

This work was supported by the National Science Foundation through a
grant for the
Institute for Theoretical Atomic and Molecular Physics at Harvard
University and
the Smithsonian Astrophysical Observatory.

\vfil\eject

\end{document}